\documentclass{article}

\usepackage{PRIMEarxiv}

\usepackage[utf8]{inputenc} 
\usepackage[T1]{fontenc}    
\usepackage{hyperref}       
\usepackage{url}            
\usepackage{booktabs}       
\usepackage{amsfonts}       
\usepackage{nicefrac}       
\usepackage{microtype}      
\usepackage{lipsum}
\usepackage{fancyhdr}       
\usepackage{graphicx}       
\graphicspath{{media/}}     

\title{FlowECG: Using Flow Matching to Create a More Efficient ECG Signal Generator
\thanks{This preprint has not undergone peer review or any post-submission improvements or corrections. The Version of Record of this contribution will be published in "Sensors, Devices and Systems – 2025" Proceedings.
}}

\author{
    Vitalii Bondar\\
    Cherkasy State Technological University \\
    Cherkasy, Ukraine \\
    \texttt{v.v.bondar.asp24@chdtu.edu.ua} \\
    \And
    Serhii Semenov\\
    University of the National Education Commission\\
    Krakow, Poland\\
    \texttt{serhii.semenov@uken.krakow.pl} \\
    \And
    Vira Babenko\\
    Cherkasy State Technological University \\
    Cherkasy, Ukraine \\
    \texttt{v.babenko@chdtu.edu.ua} \\
    \And
    Dmytro Holovniak\\
    State Scientific Research Institute of \\
    Armament and Military Equipment Testing and Certification \\
    Cherkasy, Ukraine \\
    \texttt{dm.holovniak@gmail.com} \\
}

\begin{document}
\maketitle

\begin{abstract}
Synthetic electrocardiogram generation serves medical AI applications requiring privacy-preserving data sharing and training dataset augmentation. Current diffusion-based methods achieve high generation quality but require hundreds of neural network evaluations during sampling, creating computational bottlenecks for clinical deployment. We propose FlowECG, a flow matching approach that adapts the SSSD-ECG architecture by replacing the iterative diffusion process with continuous flow dynamics. Flow matching learns direct transport paths from noise to data distributions through ordinary differential equation solving. We evaluate our method on the PTB-XL dataset using Dynamic Time Warping, Wasserstein distance, Maximum Mean Discrepancy, and spectral similarity metrics. FlowECG matches SSSD-ECG performance at 200 neural function evaluations, outperforming the baseline on three metrics. The key finding shows that FlowECG maintains generation quality with substantially fewer sampling steps, achieving comparable results with 10-25 evaluations compared to 200 for diffusion methods. This efficiency improvement reduces computational requirements by an order of magnitude while preserving physiologically realistic 12-lead ECG characteristics. The approach enables practical deployment in resource-limited clinical settings where real-time generation or large-scale synthetic data creation is needed.
\end{abstract}


\section{Introduction}

Synthetic electrocardiogram generation has become a significant research area in medical AI, driven by the growing need for privacy-preserving data sharing, training dataset augmentation, and clinical decision support systems. Modern deep generative models,
as sophisticated probability transformation functions, have shown considerable promise in capturing complex patterns \cite{bondar2025deep}. However, current state-of-the-art approaches face significant computational efficiency challenges that limit their practical deployment.

Early ECG generation relied primarily on mathematical modeling approaches, utilizing physiological simulations and established transformations, such as the Dower method \cite{quiroz2019generation}. While these methods provided good physiological accuracy, they struggled to capture the full range of pathological variations present in clinical data. The development of machine learning approaches marked a significant shift toward data-driven generation methods.

Generative Adversarial Networks represented the first breakthrough in neural ECG synthesis. The WaveGAN architecture, originally developed for audio generation, was successfully adapted for ECG data \cite{donahue2018adversarial}. Building on this foundation, Pulse2Pulse demonstrated the feasibility of generating 10-second 12-lead ECG recordings while
preserving critical clinical features like QT/RR interval relationships \cite{thambawita2021deepfake}. Several specialized GAN variants emerged, including SynSigGAN for biomedical signal generation \cite{hazra2020synsiggan} and transformer-enhanced approaches that achieved improved classification accuracy \cite{xia2023generative}. Despite these advances, GAN-based methods often suffer from training instability and mode collapse issues.

The current state-of-the-art is dominated by diffusion-based approaches, which have largely addressed the stability problems of GANs. SSSD-ECG established a new benchmark by integrating structured state space models with denoising diffusion probabilistic
models, enabling effective conditioning on multiple diagnostic labels from the PTB-XL dataset \cite{alcaraz2023diffusion}. This approach demonstrated superior performance in both quantitative metrics and clinical validation studies. Several extensions have followed, including DiffECG, which provides a versatile framework for generation, imputation, and forecasting tasks \cite{diffecg}. The DSAT-ECG architecture further improved upon these results by incorporating State Space Augmented Transformers \cite{zama2023ecg}, while BioDiffusion extended the
approach to general biomedical signal synthesis \cite{li2024biodiffusion}.

Less developed alternative approaches are variational autocoders, which offer better interpretability, with latent dimensions that can correspond to clinically relevant ECG components \cite{zhao2024vaeeg}. However, these methods typically produce over-smoothed signals that lack the fine-grained details present in real ECGs. Simulator-based approaches like SimGANs have demonstrated effectiveness in specific scenarios \cite{golany2020simgans}, but their applicability remains limited to well-understood cardiac conditions.

Despite these advances, current diffusion-based methods face a critical limitation: computational efficiency. The iterative denoising process requires hundreds of forward passes through the neural network during generation, creating bottlenecks for real-time
applications and large-scale synthetic dataset creation that limits the practical deployment of these methods in clinical environments.

The theoretical advantages of flow matching include more stable training dynamics and significantly reduced sampling time compared to iterative diffusion approaches. This approach has demonstrated success in related domains, with PeriodWave achieving high-quality waveform generation through period-aware flow matching \cite{lee2024periodwave}.

Current approaches of evaluation of synthetic ECG quality combine traditional signal processing metrics with more sophisticated measures designed for generative models. The PTB-XL dataset has emerged as the primary benchmark, providing over 21,000 clinical 12-lead ECG recordings with comprehensive diagnostic annotations \cite{wagner2020ptb}.

However, the field still lacks standardized evaluation protocols that adequately balance statistical fidelity, clinical relevance, and computational efficiency considerations.

Our work addresses the efficiency limitations of current ECG generation methods by proposing FlowECG, a flow matching approach that adapts the proven SSSD-ECG architecture while replacing the computationally expensive diffusion process with efficient flow dynamics. Our key contribution is demonstrating that flow matching can achieve comparable generation quality to state-of-the-art diffusion models while requiring significantly fewer neural function evaluations during sampling.

\section{Methods}

\subsection{Dataset}

For our experiments, we used the PTB-XL dataset, one of the largest publicly available collections of clinical 12-lead ECG recordings. This dataset contains 21,799 recordings from 18,869 patients. Each sample captures 10 seconds of cardiac activity. The patient demographics are balanced. The dataset comprises 52\% male and 48\% female participants, with the age range from 0 to 95 years and a median age of 62 years \cite{wagner2020ptb}.

We followed the dataset splitting strategy: the training portion was used to develop our generative model, and the hold-out test set was used only for the final evaluation and models comparison.

We focus on conditional ECG generation, where we use the multi-label diagnostic codes only as a conditioning rather than supervision targets. This allows us to generate new ECG signals consistent with specified medical conditions.

\subsection{Flow matching}

Flow matching represents a simulation-free approach to continuous normalizing flows
that learns a time-dependent velocity field \(u_t(x)\) for transporting samples from a base distribution \(p_0\) (such as Gaussian noise) to the target data distribution \(p_1\). This transport occurs through solving an ordinary differential equation (ODE).
A neural network \(v_{\theta}(x, t)\)) parameterizes the velocity field to match this target velocity. The training objective minimizes the mean-squared error:
\begin{equation}
    L_{FM}(\theta)=\mathbb{E}_{t{\sim}U(0,1),x{\sim}p_t} || v_\theta(x,t) - u_t(x) ||^2 .
\end{equation}

Sample generation proceeds by solving the learned ODE starting from an initial condition \(x(0) \sim p_0\):
\begin{equation}
    \frac{dx}{dt} = v_\theta(x,t)
\end{equation}

Deterministic integration from \(t=0\) to \(t=1\) produces samples \(x(1) \sim p_1\) from the target distribution.

This framework leverages linear velocity fields in latent space combined with a direct regression-based training objective. The approach avoids the computational overhead of likelihood or score function simulation while establishing a clear deterministic mapping between noise and data through ODE-based flow dynamics \cite{lipman2022flow}.

\subsection{Metrics}
We evaluate the fidelity of generated ECG signals using four complementary metrics: Dynamic Time Warping (DTW), Wasserstein distance, Mean Maximum Discrepancy, and a spectral similarity score. These metrics are commonly adopted in prior deep ECG generation studies \cite{diffecg}.

We follow standard practice by computing each metric per channel and reporting the average across all leads.

\subsubsection{Dynamic Time Warping (DTW)}
DTW aligns two signals by allowing non-linear time warping to minimize the cumulative distance. For two sequences \(x=(x_1,x_2,\dots,x_n)\) and \(y=(y_1,y_2,\dots,y_n)\) , let the cost matrix \(D(n, m)\) be:
\begin{equation}
    D(n,m) = \min \{ D(n-1,m), D(n,m-1), D(n-1,m-1) \} + d(x_n,y_n),
\end{equation}
with boundary \(D(1,1) = d(x_1, y_1)\), and the final DTW distance \(DTW(x,y)=D(N,M)\). Typically \(d(x,y)\) is the squared Euclidean or absolute \(L1\) distance.

\subsubsection{Wasserstein Distance}
The \(p\)-Wasserstein distance compares two distributions. For empirical distributions \((\alpha,\beta)\), it is defined as:
\begin{equation}
    W_p(\alpha,\beta) = (\inf_{\gamma \in \Gamma} {\mathbb{E}_{(x,y)}[|| x-y ||^p] })^{1/p} ,
\end{equation}
where \(\Gamma\) denotes all couplings of \(\alpha\) and \(\beta\).

Our implementation first extracts comprehensive features from multi-channel ECG data. For each channel c and sample i, we compute a feature vector containing various statistical moments, signal characteristics, temporal features, and spectral features.
Then all the features are combined, and 1-Wasserstein distance between the real and generated feature distributions is computed.

\subsubsection{Spectral Similarity Score}
We employ a spectral similarity score to evaluate the frequency content preservation between real and synthetic ECG signals. This metric assesses whether essential spectral characteristics, including peaks associated with heartbeats, QRS complexes, and T-waves, are adequately maintained in the generated samples.

Spectral similarity score is calculated as the normalized reverse Wasserstein distance between real and generated power spectral density features:

\begin{equation}
    Spec\ sim = \frac{1}{1+\frac{1}{n}\sum W(PSD_{real},PSD_{gen})}.
\end{equation}

Spectral similarity metric converts frequency-domain discrepancies into a single, easily interpretable score that reflects how effectively the generative model maintains clinically important oscillatory patterns \cite{wang2015specsim}.

\subsubsection{Maximum Mean Discrepancy}

The Maximum Mean Discrepancy (MMD) is a kernel-based metric that measures the difference between two distributions by comparing their mean embeddings in a reproducing kernel Hilbert space (RKHS) \cite{gretton2012kernel}. It is widely used to evaluate generative models due to its simplicity and theoretical guarantees \cite{cui2020calibrated}.

With a positive-definite kernel \(k\) and associated feature map \(\phi\), the squared MMD between the two distributions \(P\) and \(Q\) is defined as:
\begin{equation}
    MMD^2(P,Q) = || \mathbb{E}_{x \sim P}[\phi(x)] - \mathbb{E}_{y \sim Q}  [\phi(y)] ||^2_{\mathcal{H}}.
\end{equation}
This can be estimated empirically using samples \(\{x_i\}^m_{i=1} \sim P\) and \(\{y_j\}^n_{j=1} \sim Q\):
\begin{equation}
    MMD^2(P,Q) = \frac{1}{m^2}\sum_{i,i'}k(x_i, x_{i'})+\frac{1}{n^2}\sum_{j,j'}k(y_j,y_{j'})-\frac{2}{mn}\sum_{i,j}k(x_i,y_j).
\end{equation}

In our ECG generation context, we let \(P_c\) be the real-data distribution and \(Q_c\) the model-generated distribution for each channel c.

\section{Experiments and Results}

\subsection{Baseline}

We employ SSSD-ECG as our primary baseline, which represents the current state-of-the-art in conditional ECG generation using diffusion models. This baseline combines structured state space models with probabilistic diffusion to generate 12-lead ECGs
conditioned on clinical diagnostic statements. The architecture builds upon the SSSDS4 framework, where traditional dilated convolutions from the DiffWave \cite{kong2020diffwave} audio synthesis model are replaced with bidirectional S4 layers. This modification enables bette capture of long-range temporal dependencies that are characteristic of ECG signals \cite{alcaraz2023diffusion}.

The model uses 36 stacked residual layers with 256 residual and skip channels. It incorporates a three-level diffusion embedding with dimensions (128, 256, 256) to process temporal information effectively. Conditional information comes from 71 binary ECG diagnostic statements derived from the PTB-XL dataset. These binary vectors are transformed into continuous representations through learnable weight matrices and then integrated into the diffusion process at multiple layers.

The baseline generates 8 independent ECG leads, specifically the 6 precordial leads plus leads I and aVF, across 1000 timesteps representing 10 seconds of ECG data. The remaining 4 limb leads are reconstructed using established electrocardiographic relationships: \(III = II-I\), \(aVL = \frac{I-III}{2}\), \(aVF = \frac{II+III}{2}\), and \(-aVR = \frac{I+II}{2}\). This approach ensures that generated ECGs maintain physiologically consistent lead relationships.

Training configuration includes 200 diffusion timesteps with a linear noise schedule where \(\beta\) ranges from 0.0001 to 0.02. The model uses Adam optimization with a learning rate of \(2 \cdot 10^{-4}\) and mean squared error loss. Alcaraz et al. demonstrated that SSSD-ECG achieves superior performance compared to conditional GAN-based approaches, including WaveGAN \cite{donahue2018adversarial} and Pulse2Pulse \cite{thambawita2021deepfake}, across multiple evaluation metrics such as classifier-based quality assessment and expert clinical evaluation. This established performance benchmark provides a solid foundation for evaluating our proposed flow matching approach in terms of generation quality and computational efficiency \cite{alcaraz2023diffusion}.

\subsection{FlowECG}

We propose FlowECG, which adapts the established SSSD-ECG architecture to use flow matching instead of diffusion for conditional ECG generation. Our method keeps the same model structure as the baseline, including the 36 stacked residual layers and bidirectional S4 components, allowing us to directly compare the two training approaches.

The main difference between our approach and SSSD-ECG lies in the formulation of the generative process. While SSSD-ECG uses discrete diffusion steps, FlowECG employs a continuous flow matching framework. We sample time uniformly \(t \in [0;1]\) during training and create a linear interpolation between the target ECG signal \(x_1\) and Gaussian noise \(x_0\).

This interpolation path connects the data distribution to a simple noise distribution. The network learns to predict the vector field \(v_t = x_1 - x_0\), which points from noise toward the true data at each interpolation point.

Our training objective uses standard \(L2\) loss between the predicted and desired vector fields:
\begin{equation}
    L = \mathbb{E} || f_\theta(x_t,c,t) - (x_1-x_0) ||^2.
\end{equation}

Here, \(f_\theta\) represents the neural network, \(c\) contains the conditional diagnostic labels, and the expectation covers data samples, noise, time steps, and conditioning information. This formulation is simpler than the weighted variational bound used in diffusion models.

Generation works by solving the ordinary differential equation 
\(\frac{dx}{dt} = f_\theta(x_t,c,t)\) from random noise to data. We start with Gaussian noise and integrate forward using the learned vector field, conditioned on the desired ECG diagnostic statements. The conditioning mechanism remains identical to SSSD-ECG, where 71 binary diagnostic labels are embedded and fed into multiple network layers.

We use the same optimization settings as the baseline for fair comparison, including
the Adam optimizer with learning rate \(2 \cdot 10^{-4}\) and batch size 6. The continuous nature of flow matching potentially offers advantages over the discrete 200-step diffusion process, particularly in terms of sampling efficiency and training dynamics.

\subsection{Results}

We conducted comprehensive experiments comparing FlowECG against the retrained SSSD-ECG baseline using identical architectural and training configurations. Both models were evaluated on their ability to generate high-quality 12-lead ECGs conditioned on diagnostic labels from the PTB-XL dataset. Our evaluation encompasses visual quality assessment, quantitative metrics analysis, and sampling efficiency under varying numbers of neural function evaluations.

\begin{figure}
    \centering
    \includegraphics[width=1.0\linewidth]{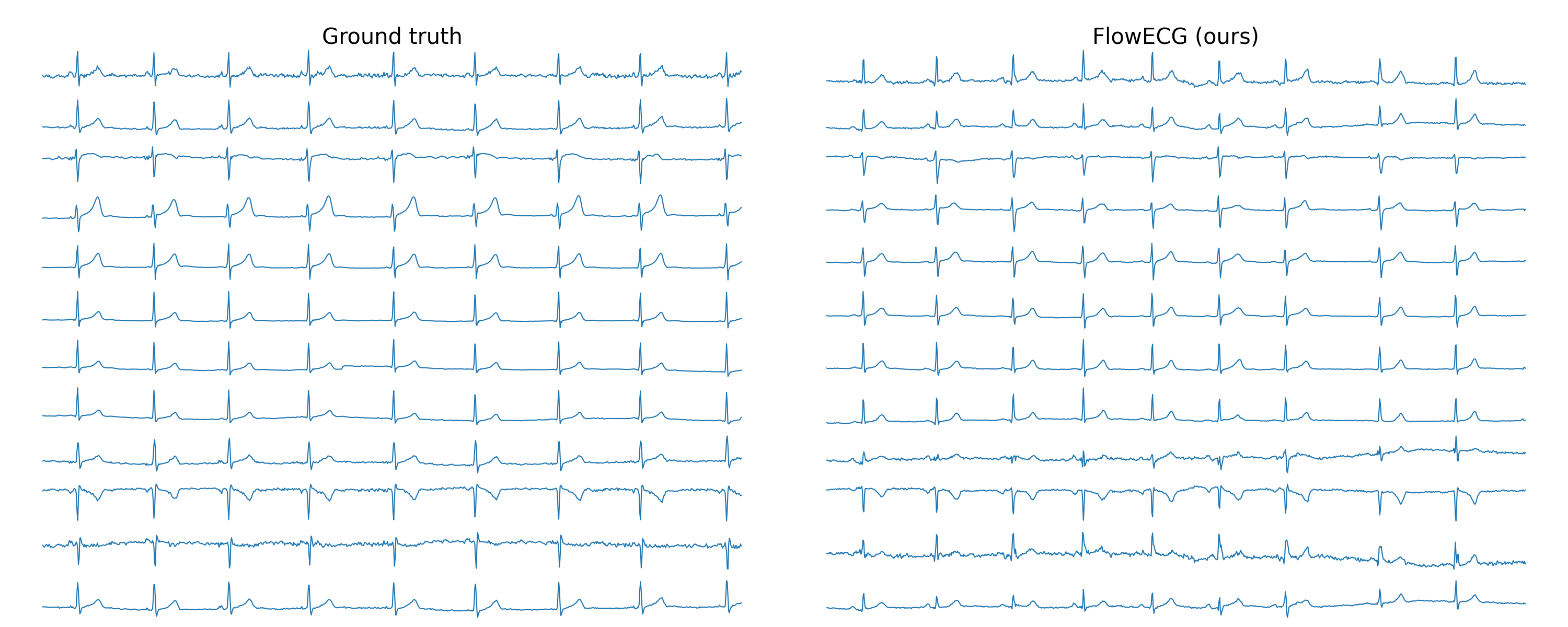}
    \caption{Visual comparison of ground truth ECGs (left) and FlowECG-generated samples (right) showing 12-lead electrocardiogram waveforms.}
    \label{fig:vis-fig}
\end{figure}

Visual comparison between ground truth ECGs and FlowECG-generated samples shows that our approach produces physiologically plausible waveforms with appropriate morphological characteristics across all 12 leads (Figure \ref{fig:vis-fig}). Generated ECGs exhibit consistent rhythm patterns, proper lead relationships, and realistic amplitude variations that closely match the original data distribution. The temporal dynamics and inter-lead correlations appear well-preserved, indicating that FlowECG successfully captures the complex dependencies in multi-lead ECG signals.

Quantitative evaluation using four established metrics confirms the competitive performance of FlowECG compared to SSSD-ECG when using 200 neural function evaluations. As described in Section 2.3, we computed Maximum Mean Discrepancy (MMD), Dynamic Time Warping (DTW) distance, Wasserstein distance, and Spectral Similarity Score for both approaches. Table \ref{tab:res-table} shows that FlowECG achieves superior performance on three out of four metrics, with notable improvements in MMD and Wasserstein distance. The DTW distance shows a modest increase, while Spectral Similarity Score slightly worsened.

\begin{table}
 \caption{Quantitative comparison of SSSD-ECG and FlowECG performance metrics using 200 neural function evaluations}
  \centering
  \begin{tabular}{lcccc}
    \toprule
     & MMD $\downarrow$ & DTW $\downarrow$ & Wasserstein $\downarrow$ & SimScore $\uparrow$\\
    \midrule
    SSSD-ECG & 80.70 & \textbf{73.49} & 1.03 & 0.30 \\
    FlowECG (ours) & \textbf{21.09} & 83.96 & \textbf{0.62} & \textbf{0.37} \\
    \bottomrule
  \end{tabular}
  \label{tab:res-table}
\end{table}

\begin{figure}
    \centering
    \includegraphics[width=1\linewidth]{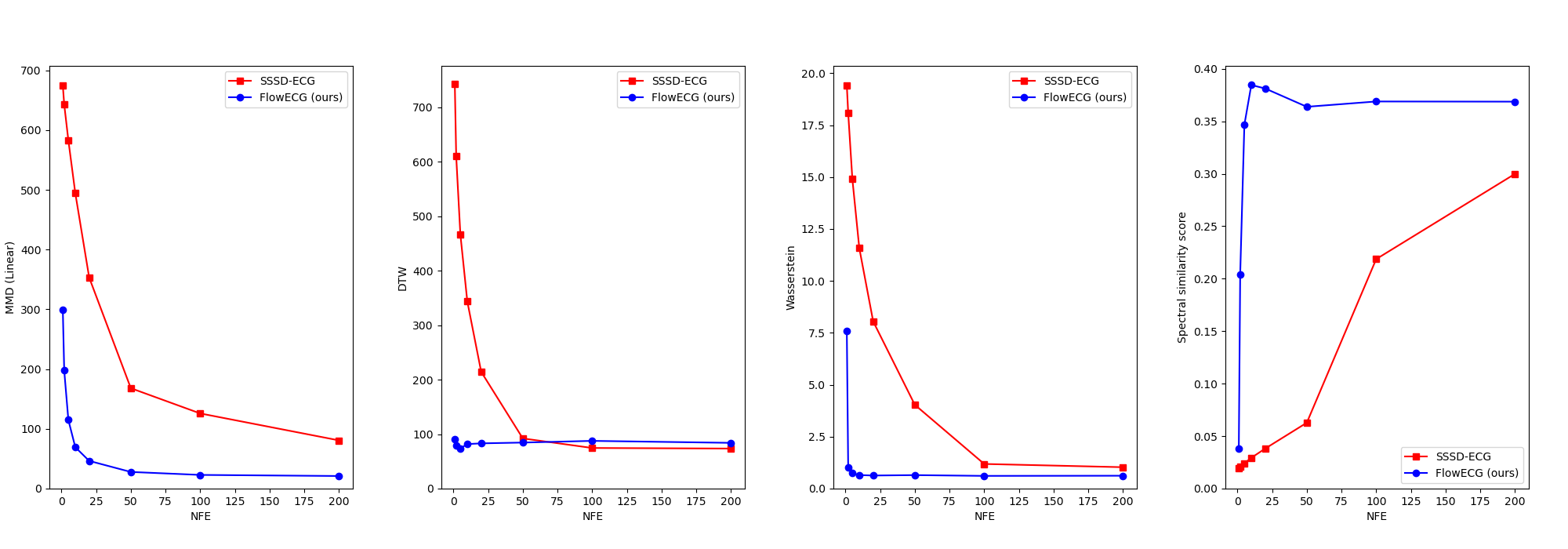}
    \caption{Performance degradation analysis showing metric values across different numbers of neural function evaluations for SSSD-ECG and FlowECG models.}
    \label{fig:fig-metrics}
\end{figure}

The most significant finding emerges when examining model performance under reduced sampling budgets. Neural Function Evaluations (NFE) represent the number of forward passes through the neural network during generation, corresponding to sampling steps in the numerical integration process. For SSSD-ECG, this relates to denoising steps in the reverse diffusion process, while for FlowECG, it corresponds to integration steps when solving the flow ordinary differential equation using the Euler method. Figure \ref{fig:fig-metrics} demonstrates that SSSD-ECG exhibits dramatic quality degradation
when reducing NFE from 200 to lower values, with DTW distance increasing to over 600 at 2 NFE and over 200 at 25 NFE. All other metrics show same dynamic of the degradation. FlowECG demonstrates superior robustness to NFE reduction, maintaining consistently high quality even with fewer sampling steps.

This sampling efficiency advantage has practical implications for clinical deployment where computational resources may be limited or real-time generation is required. The ability to generate quality ECGs with 10-20 NFE instead of 200 represents a 10-20x reduction in computational cost, as each NFE reduction directly translates to fewer neural network forward passes. This efficiency gain becomes valuable in applications requiring batch generation of synthetic ECGs for data augmentation, privacy-preserving data sharing, or real-time clinical systems.

The consistent performance of FlowECG across different NFE budgets suggests that the flow matching framework provides a more stable optimization landscape compared to the discrete diffusion process. While SSSD-ECG requires careful scheduling of noise levels across many timesteps, FlowECG's continuous formulation learns smoother vector fields that can be integrated accurately with fewer steps. This property makes FlowECG suitable for scenarios where computational budget may vary or where adaptive sampling strategies could be employed based on conditioning complexity.

\section{Conclusions}

We presented FlowECG, a flow matching approach for conditional ECG generation that adapts the established SSSD-ECG architecture to a continuous generative framework. Our method demonstrates that flow matching can achieve competitive generation quality while offering significant computational advantages over diffusion-based approaches for medical time series synthesis.

Experimental results show that FlowECG maintains high-quality generation capabilities while providing superior sampling efficiency. At 200 neural function evaluations, FlowECG outperforms the baseline on three out of four quantitative metrics, with notable improvements in Maximum Mean Discrepancy and Wasserstein distance. Generated ECGs exhibit physiologically realistic morphology and preserve important clinical characteristics across all 12 leads.

The key finding reveals that flow matching enables effective ECG generation with dramatically fewer sampling steps than diffusion models. While SSSD-ECG requires the full 200-step process to maintain quality, FlowECG produces comparable results with 10-25 neural function evaluations. This computational reduction makes the approach practical for real-time clinical applications and deployment in resource-constrained environments.

Our analysis indicates that the continuous formulation provides a more stable optimization landscape than discrete diffusion processes. The learned vector fields integrate accurately with fewer steps, eliminating the need for careful noise scheduling across many timesteps. This stability proves particularly valuable when computational budgets vary or when adaptive sampling becomes necessary.

These efficiency gains enable practical deployment across various clinical scenarios, including data augmentation for ECG analysis models, privacy-preserving synthetic data sharing, and real-time generation for clinical decision support. The reduced computational requirements address a significant barrier to implementing generative models in clinical settings where processing power may be limited.

\bibliographystyle{unsrt}  
\bibliography{references}

\begin{thebibliography}{10}

\bibitem{bondar2025deep}
Vitalii Bondar, Vira Babenko, Roman Trembovetskyi, Yurii Korobeinyk, and Viktoriya Dzyuba.
\newblock Deep generative models as the probability transformation functions.
\newblock {\em arXiv preprint arXiv:2506.17171}, 2025.

\bibitem{quiroz2019generation}
MA~Quiroz-Ju{\'a}rez, O~Jim{\'e}nez-Ram{\'\i}rez, R~V{\'a}zquez-Medina, V~Bre{\~n}a-Medina, JL~Arag{\'o}n, and RA~Barrio.
\newblock Generation of ecg signals from a reaction-diffusion model spatially discretized.
\newblock {\em Scientific reports}, 9(1):19000, 2019.

\bibitem{donahue2018adversarial}
Chris Donahue, Julian McAuley, and Miller Puckette.
\newblock Adversarial audio synthesis.
\newblock {\em arXiv preprint arXiv:1802.04208}, 2018.

\bibitem{thambawita2021deepfake}
Vajira Thambawita, Jonas~L Isaksen, Steven~A Hicks, Jonas Ghouse, Gustav Ahlberg, Allan Linneberg, Niels Grarup, Christina Ellervik, Morten~Salling Olesen, Torben Hansen, et~al.
\newblock Deepfake electrocardiograms using generative adversarial networks are the beginning of the end for privacy issues in medicine.
\newblock {\em Scientific reports}, 11(1):21896, 2021.

\bibitem{hazra2020synsiggan}
Debapriya Hazra and Yung-Cheol Byun.
\newblock Synsiggan: Generative adversarial networks for synthetic biomedical signal generation.
\newblock {\em Biology}, 9(12):441, 2020.

\bibitem{xia2023generative}
Yi~Xia, Yangyang Xu, Peng Chen, Jun Zhang, and Yongliang Zhang.
\newblock Generative adversarial network with transformer generator for boosting ecg classification.
\newblock {\em Biomedical Signal Processing and Control}, 80:104276, 2023.

\bibitem{alcaraz2023diffusion}
Juan Miguel~Lopez Alcaraz and Nils Strodthoff.
\newblock Diffusion-based conditional ecg generation with structured state space models.
\newblock {\em Computers in biology and medicine}, 163:107115, 2023.

\bibitem{diffecg}
Nour Neifar, Achraf Ben-Hamadou, Afef Mdhaffar, and Mohamed Jmaiel.
\newblock Diffecg: A versatile probabilistic diffusion model for ecg signals synthesis.
\newblock In {\em 2024 IEEE/ACIS 22nd International Conference on Software Engineering Research, Management and Applications (SERA)}, pages 182--188, 2024.

\bibitem{zama2023ecg}
Md~Haider Zama and Friedhelm Schwenker.
\newblock Ecg synthesis via diffusion-based state space augmented transformer.
\newblock {\em Sensors}, 23(19):8328, 2023.

\bibitem{li2024biodiffusion}
Xiaomin Li, Mykhailo Sakevych, Gentry Atkinson, and Vangelis Metsis.
\newblock Biodiffusion: A versatile diffusion model for biomedical signal synthesis.
\newblock {\em Bioengineering}, 11(4):299, 2024.

\bibitem{zhao2024vaeeg}
Tong Zhao, Yi~Cui, Taoyun Ji, Jiejian Luo, Wenling Li, Jun Jiang, Zaifen Gao, Wenguang Hu, Yuxiang Yan, Yuwu Jiang, et~al.
\newblock Vaeeg: Variational auto-encoder for extracting eeg representation.
\newblock {\em NeuroImage}, 304:120946, 2024.

\bibitem{golany2020simgans}
Tomer Golany, Kira Radinsky, and Daniel Freedman.
\newblock Simgans: Simulator-based generative adversarial networks for ecg synthesis to improve deep ecg classification.
\newblock In {\em International Conference on Machine Learning}, pages 3597--3606. PMLR, 2020.

\bibitem{lee2024periodwave}
Sang-Hoon Lee, Ha-Yeong Choi, and Seong-Whan Lee.
\newblock Periodwave: Multi-period flow matching for high-fidelity waveform generation.
\newblock {\em arXiv preprint arXiv:2408.07547}, 2024.

\bibitem{wagner2020ptb}
Patrick Wagner, Nils Strodthoff, Ralf-Dieter Bousseljot, Dieter Kreiseler, Fatima~I Lunze, Wojciech Samek, and Tobias Schaeffter.
\newblock Ptb-xl, a large publicly available electrocardiography dataset.
\newblock {\em Scientific data}, 7(1):1--15, 2020.

\bibitem{lipman2022flow}
Yaron Lipman, Ricky~TQ Chen, Heli Ben-Hamu, Maximilian Nickel, and Matt Le.
\newblock Flow matching for generative modeling.
\newblock {\em arXiv preprint arXiv:2210.02747}, 2022.

\bibitem{wang2015specsim}
Ke~Wang, Bin Yong, Xingfa Gu, Pengfeng Xiao, and Xueliang Zhang.
\newblock Spectral similarity measure using frequency spectrum for hyperspectral image classification.
\newblock {\em IEEE Geoscience and Remote Sensing Letters}, 12(1):130--134, 2015.

\bibitem{gretton2012kernel}
Arthur Gretton, Karsten~M. Borgwardt, Malte~J. Rasch, Bernhard Sch\"{o}lkopf, and Alexander Smola.
\newblock A kernel two-sample test.
\newblock {\em J. Mach. Learn. Res.}, 13(null):723–773, March 2012.

\bibitem{cui2020calibrated}
Peng Cui, Wenbo Hu, and Jun Zhu.
\newblock Calibrated reliable regression using maximum mean discrepancy.
\newblock {\em Advances in Neural Information Processing Systems}, 33:17164--17175, 2020.

\bibitem{kong2020diffwave}
Zhifeng Kong, Wei Ping, Jiaji Huang, Kexin Zhao, and Bryan Catanzaro.
\newblock Diffwave: A versatile diffusion model for audio synthesis.
\newblock {\em arXiv preprint arXiv:2009.09761}, 2020.

\end{thebibliography}

\end{document}